\relax

\documentclass[letterpaper]{article} 
\usepackage{aaai21}  
\usepackage{times}  
\usepackage{helvet} 
\usepackage{courier}  
\usepackage[hyphens]{url}  
\usepackage{graphicx} 
\usepackage{subfigure}

\usepackage{natbib}  
\usepackage{caption} 
\usepackage{natbib}  
\usepackage{caption} 
\title{Are Top School Students More Critical of Their Professors? \\
Mining Comments on RateMyProfessor.com
}
\author{
    Ziqi Tang, Yutong Wang, Jiebo Luo\\
    {University of Rochester}\\
    {ztang14@ur.rochester.edu, ywang343@ur.rochester.edu, jluo@cs.rochester.edu}\\
}

\begin{document}
\maketitle

\section*{\textbf{abstract}}
Student reviews and comments on RateMyProfessor.com reflect realistic learning experiences of  students. Such information provides a large-scale data source to examine the teaching quality of the lecturers. In this paper, we propose an in-depth analysis of these comments. First, we partition our data into different comparison groups. Next, we perform exploratory data analysis to delve into the data. Furthermore, we employ Latent Dirichlet Allocation and sentiment analysis to extract topics and understand the sentiments associated with the comments. We uncover interesting insights about the characteristics of both college students and professors. Our study proves that student reviews and comments contain crucial information and can serve as essential references for enrollment in  courses and universities.

\section*{\textbf{Introduction}}
Since 1983, the U.S. News \& World Report has been publishing rankings for the colleges and universities in the United States each fall. These rankings have remarkable impacts on applications, admissions, enrollment decisions, as well as tuition pricing policies~\cite{NBERw7227}. It is an important reference for not only students and parents, but also institutions and professors. The ranking methodology measures and calculates a variety of factors, and has been continuously refined over time based on user feedback, discussions with institutions and education experts, literature reviews and their own data~\cite{morse_brooks_2020}. The current ranking methodology considers the following factors, along with indicator weights: Graduation and Retention Rates (22\%), Undergraduate Academic (20\%), Faculty resources (20\%), Financial Resources (10\%), student selectivity for entering class (7\%), Graduation Rate performance (8\%), Social Mobility (5\%), Graduate Indebtedness (5\%), and Alumni Giving Rate (3\%). This measurement takes a good number of objective factors into consideration. However, the learning experiences of students are subjective and personal, which cannot  readily be represented by the ranking scores. In this regard, a  professor rating website such as RateMyProfessors.com is a great resource to uncover the hidden knowledge about the learning experience that the U.S. News Rankings can not account for.

Rate My Professor is a website that allows students to anonymously rate their professors and write comments. The website claims that users have added more than 19 million ratings, 1.7 million professors, and over 7,500 schools to the website, and there are more than 4 million college students using this website each month~\cite{RateMyProfessors}. Such massive text data is a great resource to study the following topics: features of different universities, learning experiences of students, and course and lecture qualities.
Past literature has primarily examined the usefulness and validity of these ratings~\cite{doi:10.1080/02602930701293405}, and the correlation levels between easiness, clarity and helpfulness of lecturers~\cite{articleA}.  Yet the rich data on Rate My Professor contain more hidden information to discover. A unique feature of Rate My Professor is that it has professor reviews from different tiers of universities, such as Ivy League schools, Big Ten schools, and community colleges. These reviews discuss the same topic, which is the experiences of taking a course from a college professor. This provides an opportunity to conduct a plausible control variable experiment to learn about the characteristics of students and professors in different universities or colleges.

In summary, this study makes several contributions:
 \begin{enumerate}
    \item We conduct a large-scale study of the course learning experiences across the broad spectrum of universities and colleges  in the United States. 
    \item We employ exploratory data analysis, topic modeling, and sentiment analysis to mine the behaviors and characteristics of different segments of colleges, students and professors. 
    \item we uncover interesting and useful insights that can be used to understand and improve the learning experience. 
\end{enumerate}

\section*{\textbf{Data Collection and Preprocessing}}
Rate My Professors data was scraped from the website. We selected about 75 universities based on the U.S. News college rankings of 2020. The rationale of our selection was the following: The eight Ivy League schools represent the top ranked private universities, ten Big Ten Academic Alliance Member universities represent the top ranked public universities, and the top 15 ranked community colleges in the United States represent the community colleges. In addition, we selected the top 25 ranked universities and those ranked in [100 - 125] in the United States.

For each university in our selections, we picked the 60 most-rated professors (not highest rated), and for each professor page, we scraped the most recent 20 comments. In total, we collected 87,436 data records, containing the following attributes: {“Professor ID”, “Professor Name”, “University”, “Department”, “Course ID”, “Quality score”, “Difficulty score”, “Comments”}. Each data record represents a review by a student on a course.

We partitioned the collected data into several datasets. The rationale was the following:
\begin{enumerate}
    \item Based on the school type, we partition the data into three categories: private (Ivy League), public (Big Ten), and community colleges. 
    \item Based on the average rating scores of the professors, we calculate the average quality score of each professor, and selected those professors with an average score above 4.0 and below 2.0 (the full score range is from 1.0 to 5.0), as the high-rating professor and low-rating professor groups, respectively.  
    \item Based on the quality score of each comment, we also create datasets for three categories: comments with a score above 4.0, comments with a scores below 2.0, and comments with a score in between. 
\end{enumerate}
In the end, we have 11 datasets for three types of comparison. Note that these datasets may overlap with each other. 

\section*{\textbf{Exploratory Data Analysis}}

We perform exploratory analysis with the following initial findings:
\begin{figure}[ht]
\includegraphics[width=8.5cm,height=5cm]{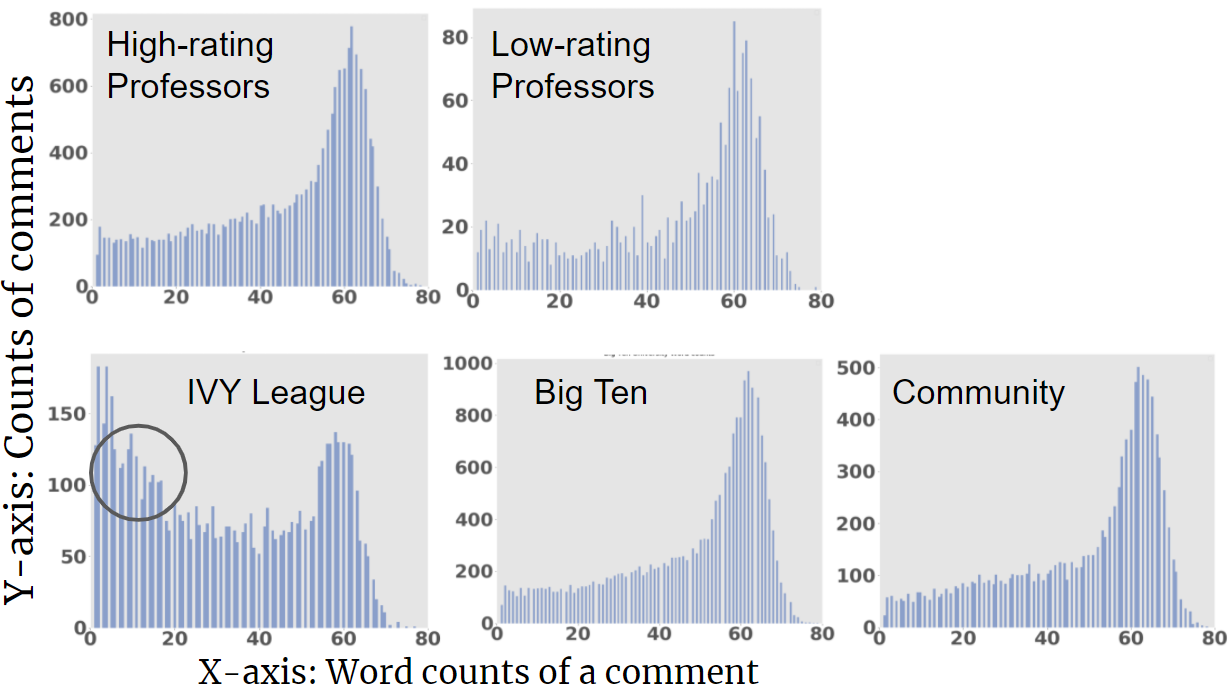}
\caption{Distributions of Word Counts in different groups.}
\label{fig:wordcounts}
\end{figure}

\begin{figure}[ht]
\includegraphics[width=8.5cm,height=5cm]{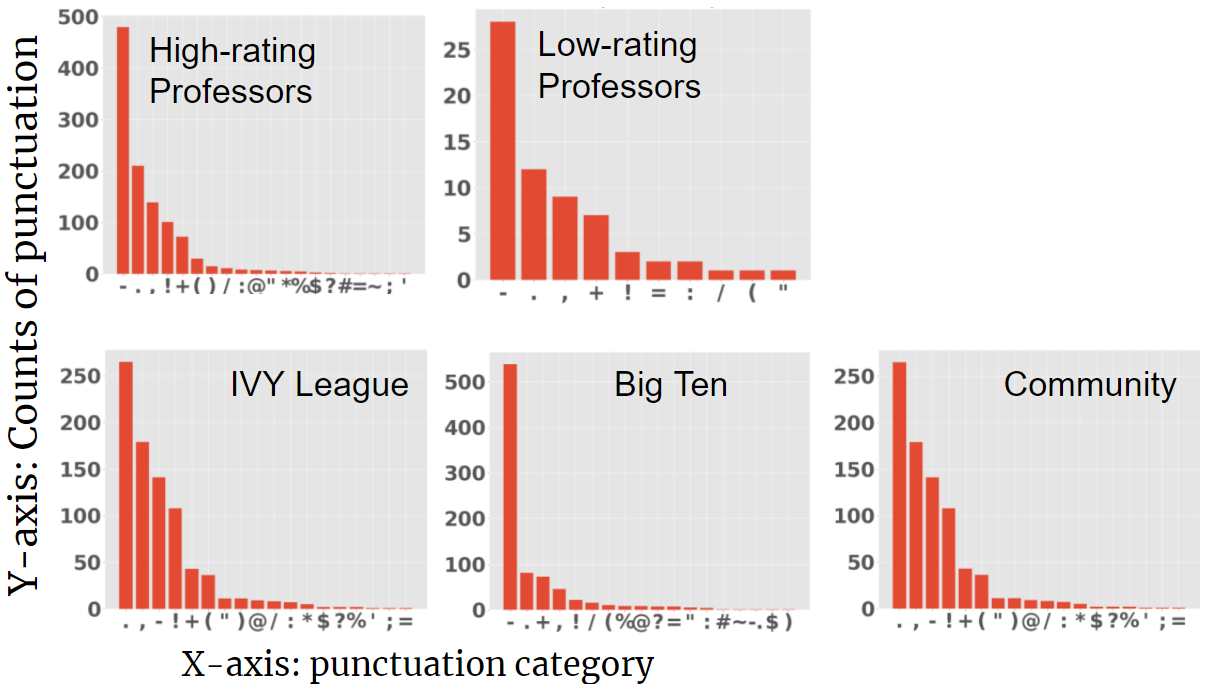}
\caption{Punctuation usage distributions.}
\label{fig:punc}
\end{figure}
\begin{figure*}[h]
\centering
\includegraphics[width=2.0\columnwidth]{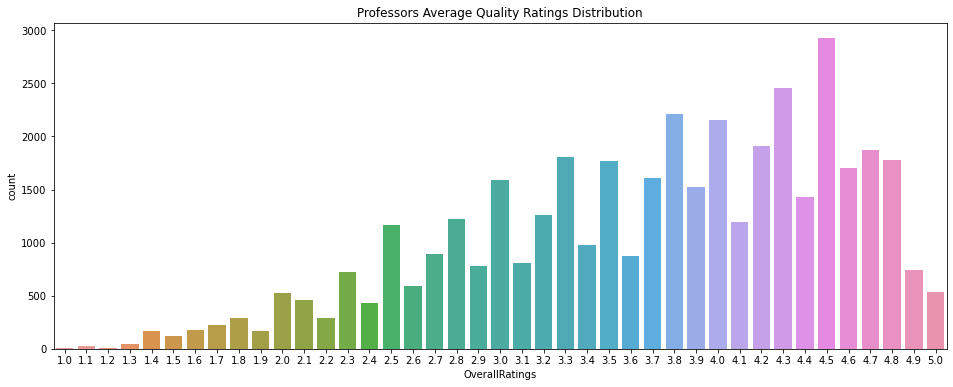}
\caption{Distribution of {\it average} quality ratings of all professors.}
\label{fig:avgrate}
\end{figure*}
\begin{figure*}[ht]
\centering
\includegraphics[width=2.0\columnwidth]{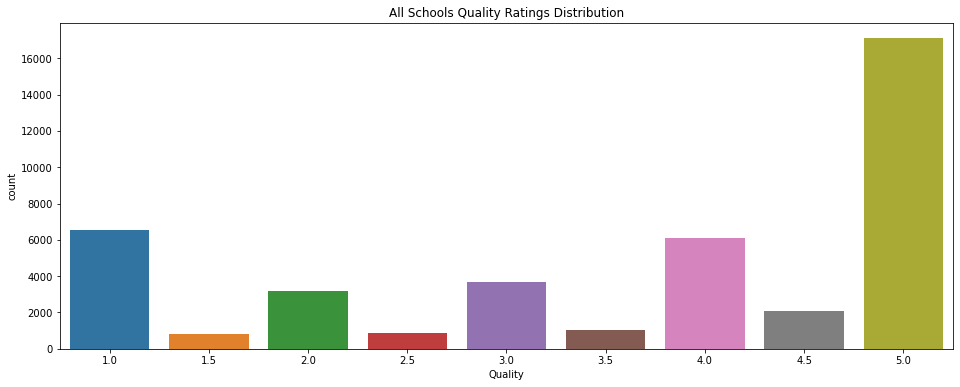}
\caption{Distribution of quality ratings of all schools.}
\label{fig:commentsrate}
\end{figure*}
\begin{figure}[ht]
\centering
\includegraphics[width=1.0\columnwidth]{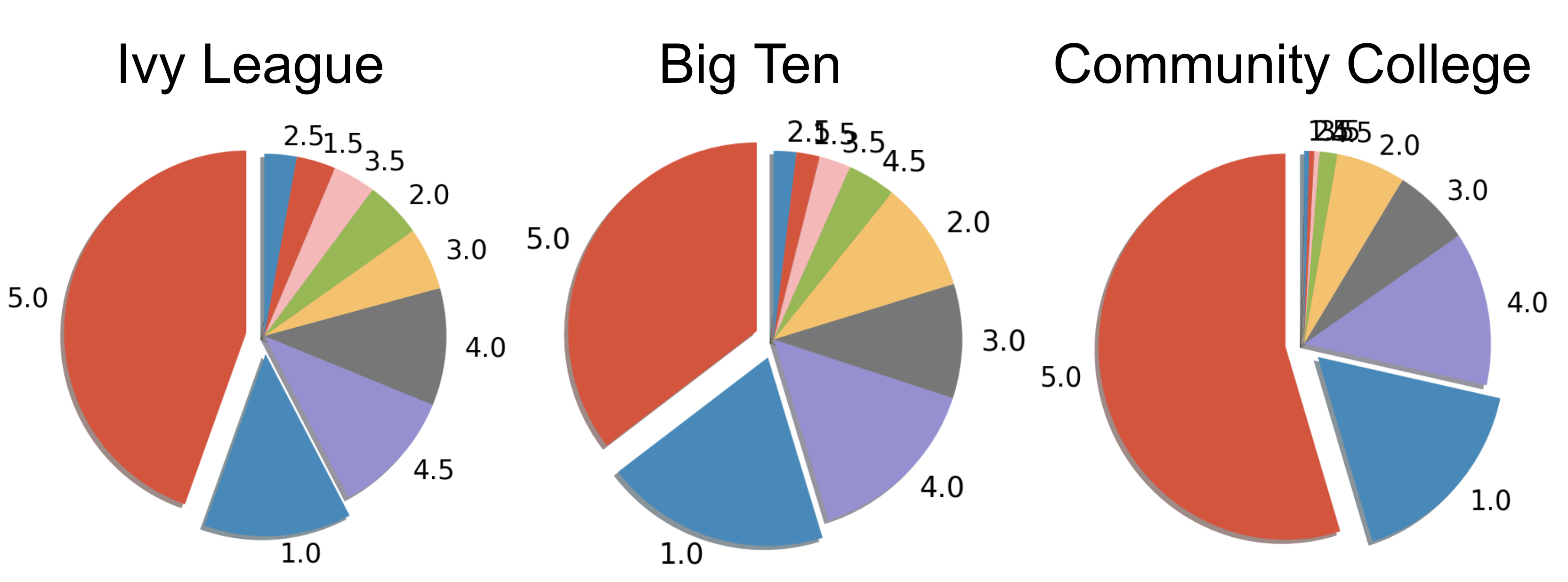}
\caption{Quality rating distributions of Ivy League, Big Ten, and community colleges.}
\label{fig:quaPie}
\end{figure}
\begin{figure}[ht]
\centering
\includegraphics[width=1.0\columnwidth]{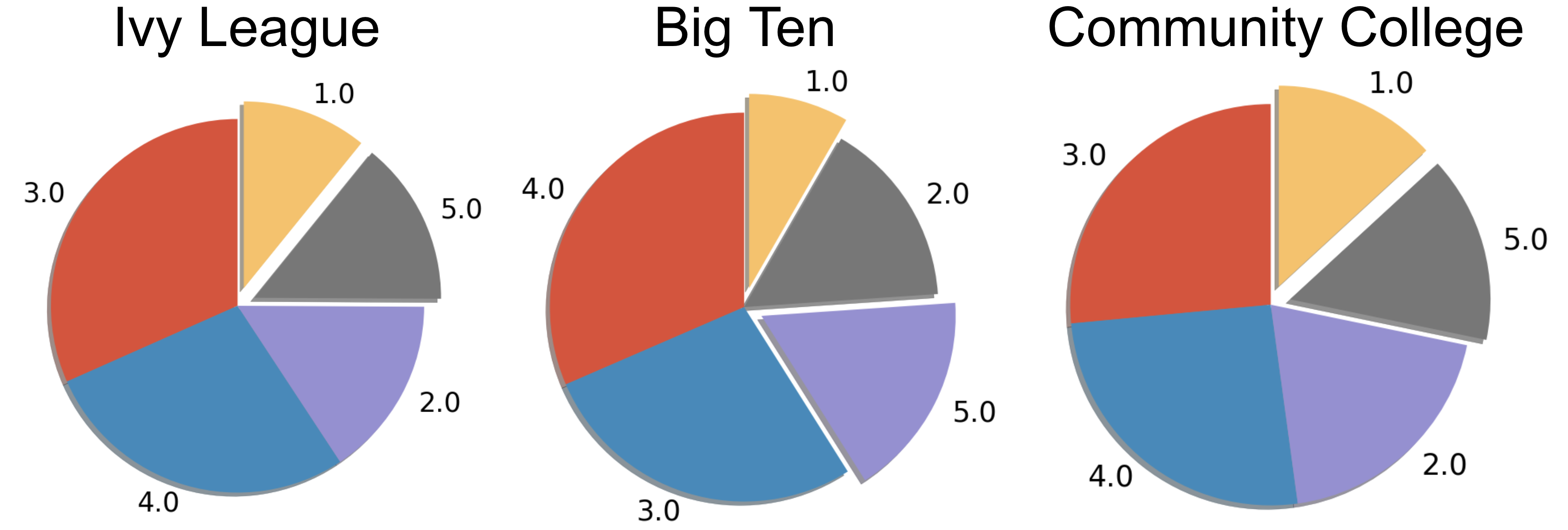}
\caption{Difficulty rating distributions of Ivy League, Big Ten, and community colleges.}
\label{fig:diffPie}
\end{figure}

\begin{itemize}
    \item The word counts shown in Figure \ref{fig:wordcounts} indicate that most of the comments contain around 60 words. All groups have similar distributions. The only difference is that Ivy League students use short phrases more often than other groups.
    \item The Punctuation usage shown in Figure \ref{fig:punc} demonstrates the most commonly used punctuation is period. The distributions for all groups are similar as well. The only difference is that community college students use commas, dashes and exclamation marks more frequently than other groups. 
    \item Figure \ref{fig:avgrate} is the distribution of average quality ratings of all professors, which has a left skewed distribution.
    \item Figure \ref{fig:commentsrate} is the distribution of quality ratings from all schools (75 schools).
    \item From Figure \ref{fig:quaPie}, the proportions of quality ratings of the three different groups of schools are different. Community college students give more high (5/5) ratings, while Ivy League students give fewer low (1/5) ratings. This answers our initial question in the title -- top school students are not critical when rating their professors and course quality.
    \item Figure \ref{fig:diffPie} shows the proportions of difficulty ratings of the three different groups of schools, which are very similar. 
    \item The correlations between quality ratings and difficulty ratings for Ivy League, Big Ten and community colleges are [-0.178, -0.424, -0.515], respectively. All groups have negative correlation values that imply the quality rating decreases when difficulty rating increases, and vice versa. Ivy League’s correlation is closer to zero which means there is little relationship between quality ratings and difficulty ratings. Moreover, students from Big Ten schools and community colleges are more likely to give a higher quality rating when the course is easy.
\end{itemize}

\section*{\textbf{Quality Rating Analysis Using Topic Modeling}}
In order to find out what factors influence the quality ratings, we perform Latent Dirichlet Allocation (LDA) to extract topics of the comments. We implement a few types of topic modeling methods: LDA, BiGram LDA and TriGram LDA using the Gensim library. Also, we apply traditional LDA and Multi-Grain LDA using the Tomotopy library. Gensim is a well-known python library for topic modeling, and Tomotopy is a new library that provides functions for topic modeling. The advantages of Tomotopy are the capability of dealing with large scale datasets, significantly faster running time than Gensim (5 to 10 times faster than Gensim), and its availability for implementing  Multi-Grain LDA. Multi-Grain LDA takes both local topics and global topics into consideration when performing topic modeling. Therefore, we decide to examine the Tomotopy Multi-Grain LDA model for our study. BiGram, TriGram and Multi-Grain LDA models are similar algorithms to the traditional LDA. However, they have an additional step that adds N-gram phrases to increase the model’s complexity, which  could be useful in boosting the model's performance. In our case, the BiGram model has phrases like: “easy-A”, “office-hour”, “online-course”, etc. For the TriGram model, there are phrases like: “extra-credit-opportunity”, “attendance\_isn\_mandatory”, etc. 

\begin{figure}[ht]
\includegraphics[width=8.5cm, height=5.5cm]{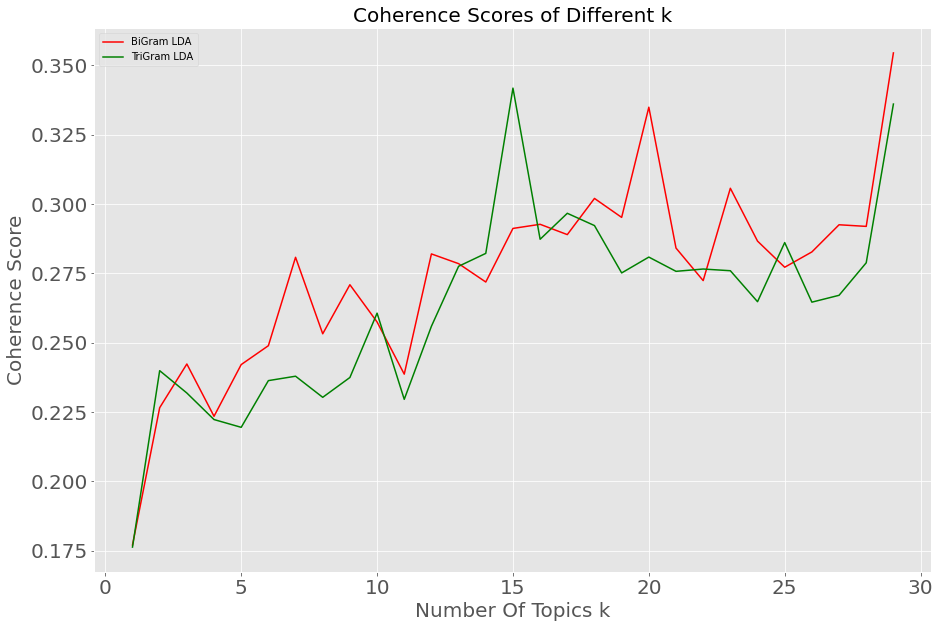}
\caption{Coherence score comparison between BiGram LDA and TriGram LDA models.}
\label{fig:coherence}
\end{figure}

\begin{figure*}[ht]
\centering
  \includegraphics[width=2.0\columnwidth]{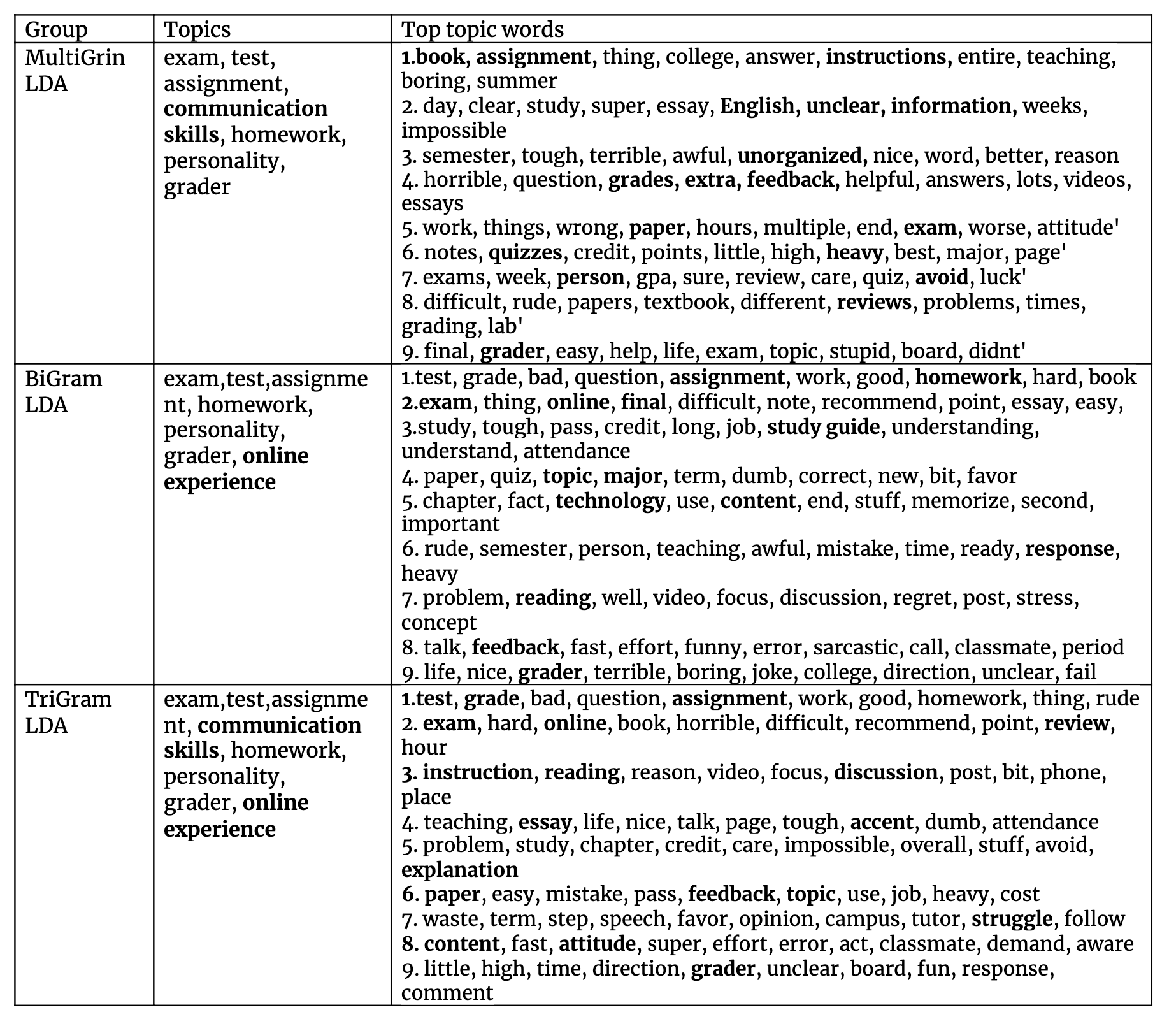}
\caption{Comparison of topic results from MultiGrin LDA, BiGram LDA and TriGram LDA models.}
\label{fig:fertableA}
\end{figure*}

In order to evaluate the performance of all of these models, we use coherence score, pyLDAvis visualization, log-likelihood, and manual checking as our evaluation metrics. For LDA, BiGram LDA and TriGram LDA models using Gensim, their coherence score comparison is shown in Figure \ref{fig:coherence}. Furthermore, we use the pyLDAvis topic modeling visualization tool to analyze the performance of models. For the Multi-Grain LDA model using Tomotopy, the library does not generate coherence scores, which is a downside of this library. Therefore, we decide to manually check all the topics these models generate and choose the one that makes more sense to us. Figure \ref{fig:fertableA} shows the resulting topics we create using BiGram LDA, TriGram LDA and Multi-Grain LDA methods. They are all generated from the same dataset (community college lower quality rating comments) and have the same number of top topics selected (nine topics). A major portion of the topics are similar. However, the TriGram LDA model covers the most of the topics. For instance, we see key word “online” from the result of TriGram LDA. Since this is a community college dataset, we can infer that community colleges tend to offer more online classes than other groups, which could be a factor that students consider when they rate the course quality. Moreover, we also see “accent” for the first time from the result of TriGram LDA. This is a very interesting factor to include because many students actually have a different time understanding their professors' accents. The communication experience is an important aspect of course quality rating.

\subsection*{Ivy League vs. Big Ten vs. Community Colleges }
\subsubsection*{Higher Ratings (4-5)}
\begin{figure*}[ht]
 \centering
 \includegraphics[width=2.0\columnwidth]{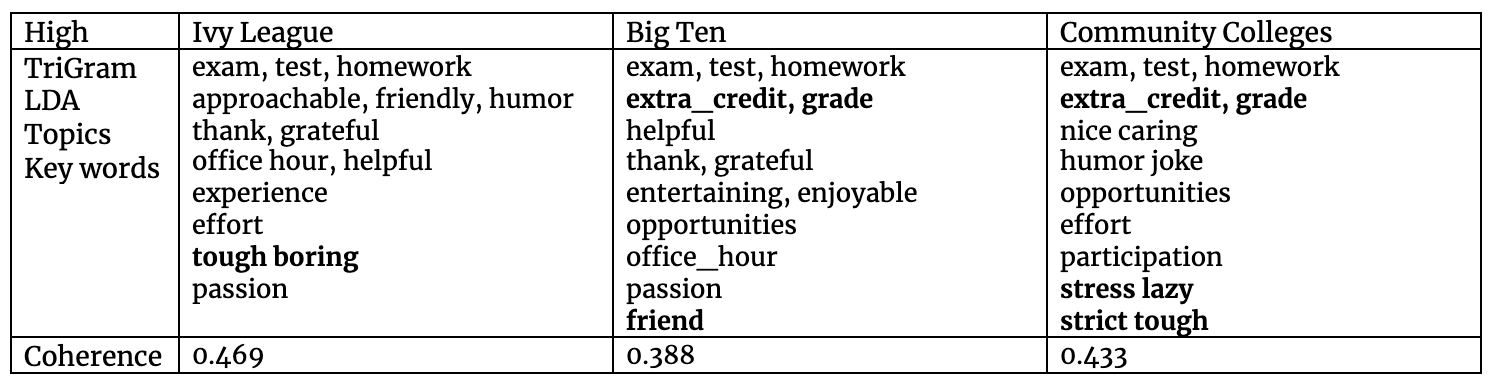}
\caption{Topic key words of higher ratings (4-5) of Ivy League vs. Big Ten vs. community colleges.}
\label{fig:fertableB}
\end{figure*}

The key words of the topics of higher quality ratings for the three groups are listed in Figure \ref{fig:fertableB}. There are many factors that students mentioned in the comments when giving higher ratings. For example, school works (homework, test, exam), extra help (office\_hour), professor’s personality (friendly, humor, entertaining), and so on. Meanwhile, some unexpected words stand out in the table: “tough”, “boring”, “strict”, implying that these are not negatively affecting Ivy League and community college’s quality ratings. In addition, both Big Ten and community college students mention “extra\_credit”, “grade” more often. The word “friend” appears in Big Ten’s topics, perhaps implying students in Big Ten schools are more likely to get along with their professors like friends. 
\subsubsection*{Lower Ratings (1-2)}
\begin{figure*}[ht]
\centering
 \includegraphics[width=2.0\columnwidth]{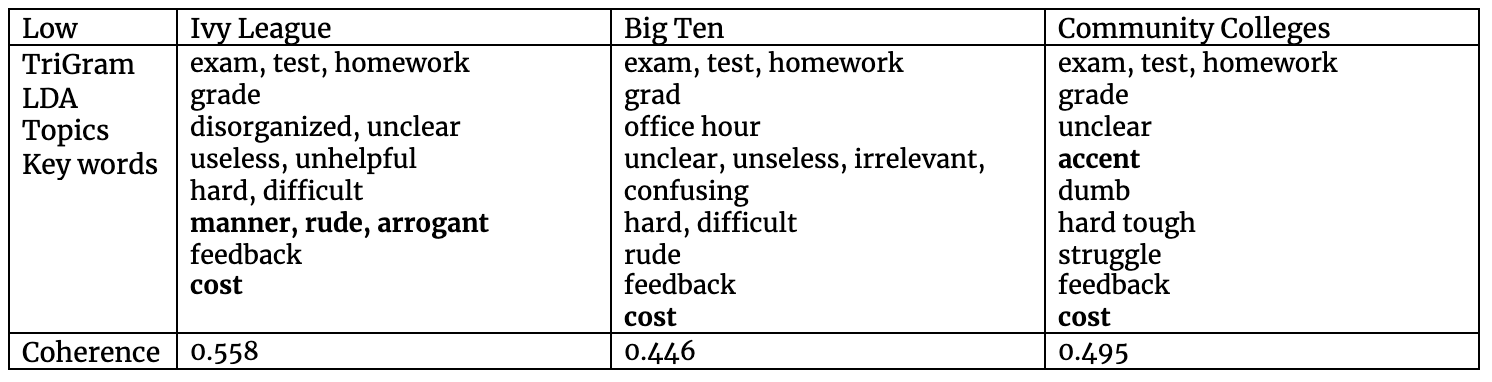}
\caption{Topic key words of lower ratings (1-2) of Ivy League vs. Big Ten vs. community colleges.}
\label{fig:fertableC}
\end{figure*}

The key words of the topics of lower quality ratings for the three groups are listed in Figure \ref{fig:fertableC}. A number of factors are mentioned by students in the comments with lower ratings. For example, school works (homework, test, exam), organization of the content (unclear, disorganized, useless), professor’s attitude (manner, rude, arrogant), and so on. One thing to point out is that “cost” is a common factor through all schools as the cost of textbooks, supplies and software has significantly negative effects on quality ratings.
\subsubsection*{Middle Ratings (2-4)}
\begin{figure*}[ht]
\centering
 \includegraphics[width=2.0\columnwidth]{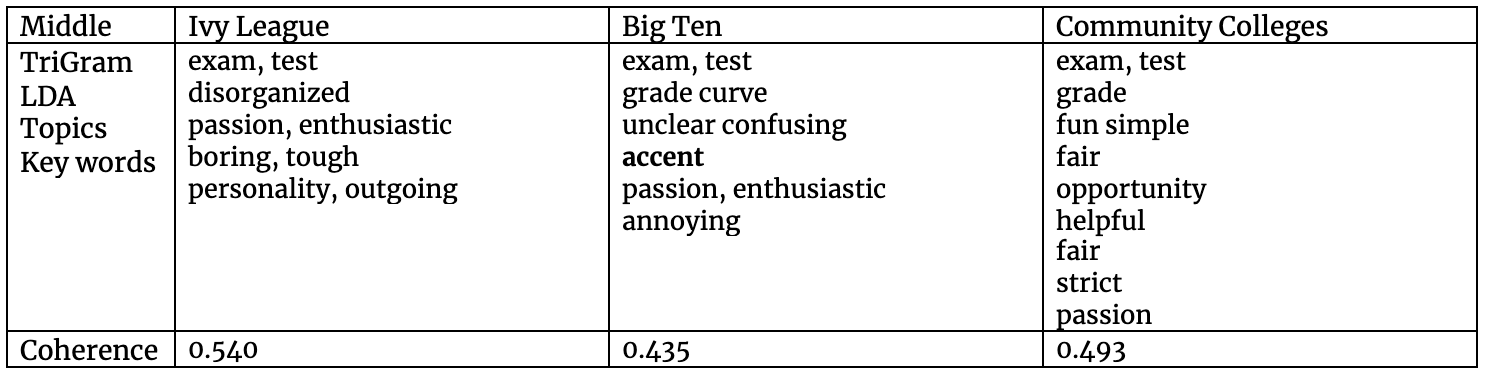}
\caption{Topic key words of middle ratings (2-4) of Ivy League vs. Big Ten vs. community colleges.}
\label{fig:fertableD}
\end{figure*}

The topic key words of middle quality ratings for the three groups are listed in Figure \ref{fig:fertableD}. The middle rating comments are usually not too extreme. We note that “accent” appears under Big Ten’s topics, and in community college’s topics for lower ratings. This suggests that Big Ten school students may have a higher tolerance for professors’ accents than community college students. 

\subsection*{High-Rating Professors vs. Low-Rating Professors}
\begin{figure*}[ht]
 \centering
 \includegraphics[width=2.0\columnwidth]{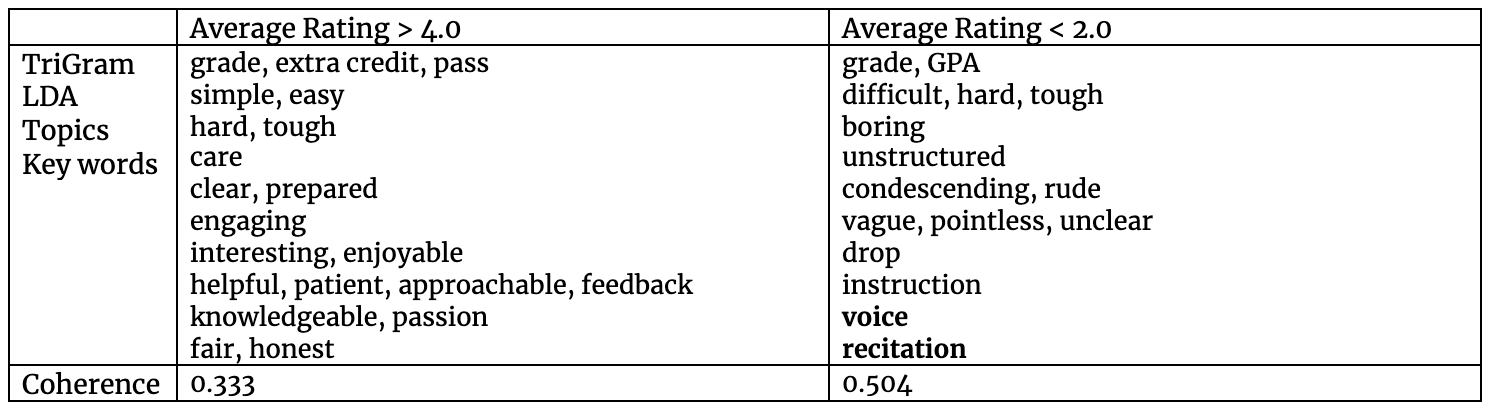}
\caption{Topic key words of professors with high average ratings vs. professors with low average ratings.}
\label{fig:fertableE}
\end{figure*}

The key words in the comments for professors with an average quality rating higher than 4 and lower than 2 are listed in Figure \ref{fig:fertableE}. One thing to notice is that the coherence score of higher rating professors is lower, which means the topics of these comments are more dissimilar. Factors that affect the average ratings of professors are: grade, difficulty, organization of the contents, personality, extra help, passion, knowledge, fairness, and so on. In contrast, “voice” and “recitation” appear in the lower rating professors category, and this is the only time they appear. This implies communication is critical to students’ experience in classes, and professors teaching science classes (Physics, Chemistry, Biology) that have recitation sections tend to get lower average ratings.

\section*{Sentiment Analysis Using A Lexical Approach}
The LIWC2015 toolkit includes the main text analysis module along with a group of predefined internal lexicons. The text analysis module compares each word in the text against the dictionary and then identifies which words are associated with which psychologically-relevant categories~\cite{pennebaker_boyd_jordan_blackburn_2015}. It has been used on previous studies for sentiment analysis on text data from social media (e.g., ~\cite{chen2020eyes}. LIWC2015 provides about a hundred psychologically-relevant categories, from which we select around 20 categories for our analysis. 

\subsection*{Ivy League vs. Big Ten vs. Community Colleges}



\begin{figure*}[ht]
\centering
\includegraphics[width=16.5cm, height=13.5cm]{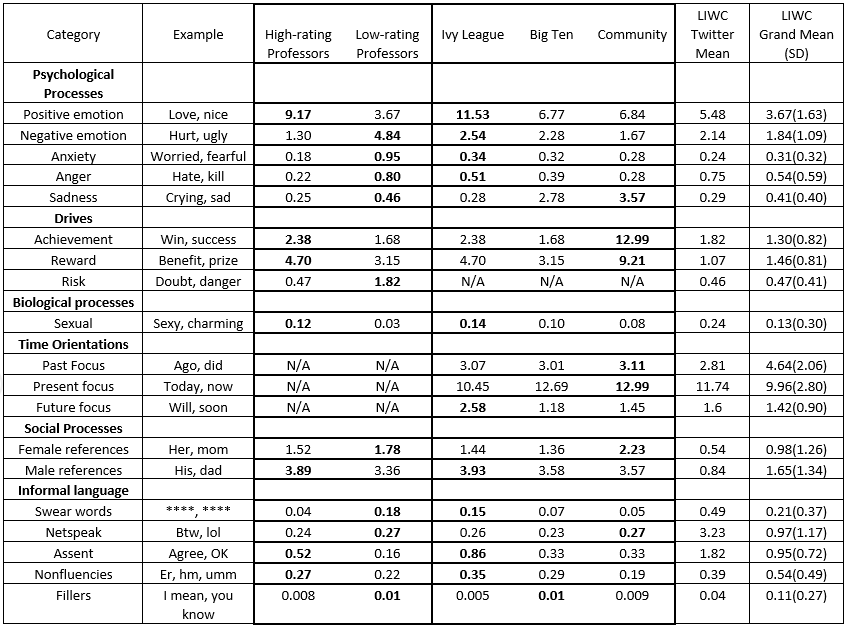}
\caption{LIWC results of two comparison groups. 
Group A: professors with average ratings above 4.0 vs. professors with average ratings below 2.0. 
Group B: Ivy League vs. Big Ten vs. community colleges. Grand Mean is the unweighted mean of the six genres, and Mean SD refers to the unweighted mean of the standard deviations across the six genre categories. }
\label{fig:ziqitable2}

\end{figure*}

After we obtain the LIWC scores for each data record, we calculate the average scores and standard deviations. Figure \ref{fig:ziqitable2} shows the LIWC results for our first comparison group (Group A). Some interesting categories stand out: positive emotion, anxiety, achievement, sexual, and gender. We run the t-test on these categories and LIWC grand average scores. The two-tailed P values for Positive Emotion, Achievement and Male Reference were all below 0.001 ($P < 0.001$). By conventional criteria, the differences are considered to be statistically significant. 

We make the following observations:
\begin{enumerate}
    \item The positive emotion scores for college students are overall higher than the average. The Ivy League students score is not only higher than the grand average, but also higher than the other groups. It indicates that students from top ranked private schools do not tend to criticize professors more, instead they praise the professors more often than other groups. 
    \item The Achievement score for community college students is higher than other groups. Our interpretation is that the community college students may have had jobs previously and they decide to attend community college because they want to receive more education and learn more skills. They possibly have clearer motivation and goals than other groups. Therefore they tend to talk about achievement-related topics more often in their comments.
    \item The Female Reference score for community colleges and Male Reference score for Ivy League schools stand out. The gender reference scores are measured when the students mention gender related phrases, such as he or she. Due to the fact that Rate My Professor website does not record the gender of the professor, we collect a fixed number of comments from each professor.  The score generated from gender reference words is the most reliable way to infer male and female professors.  Our analysis indicates that there are more male professors in Ivy league schools and more female lecturers in the community colleges. Our interpretation is that for research-oriented institutions, like Ivy League schools and Big Ten schools, the professors are required to perform research and teaching full-time. Community colleges, on the other hand, have more part-time lecturer positions. For female professors who might have to take care of family and children at the same time, teaching part-time at community college seems to be a good option.  
    \item The anxiety scores are considered to be statistically insignificant. Based on the  literature, our expectation was that students attending top ranked private colleges have a higher likelihood to feel depression and pressure~\cite{deresiewicz_2014}. However, the LIWC results show that the students did not express pressure and anxiety in their reviews. Our interpretation is that these comments were mostly written after the final exams or projects. The students no longer feel anxious at the time they post the comments. 
    \item The sexual scores are considered to be statistically insignificant. The sexual category contains phrases that describe the appearance of the professors. This could indicate whether the appearance could affect the student ratings and comments. Our study showed there is no evidence to prove the existence of connection between appearance and student ratings.
\end{enumerate}


\subsection*{High-Rating Professors vs. Low-Rating Professors}
Similarly, after we obtain the LIWC scores for each data record, we calculated the average scores and standard deviations. Figure \ref{fig:ziqitable2} also shows the LIWC results for our second comparison group (Group B). Some interesting categories stand out: Achievement and Gender. We run the t-test on these categories and LIWC grand average scores. The two-tailed P value for Achievement and Gender Reference were both less than 0.001 ($P < 0.01$). By conventional criteria, the differences were considered to be statistically significant. 

The specific findings are:
\begin{enumerate}
    \item The Achievement score for high-rating professors is higher than the low-rating professors. This may indicate that apart from the general impressions people have for a good professor, students think a good professor also needs to know how to motivate the students.
    \item The Female Reference score for low-rating professors is higher, while the Male Reference score for high-rating professors is higher.  This shows that there are more low-rating female professors and more high-rating male professors. It may imply that students are more critical on female professors than male professors.
\end{enumerate}
\section*{\textbf{Conclusion and Future Work}}
In this paper, we have presented a framework of evaluating the learning experiences of college students from a more subjective perspective. We first partition the scraped data from RateMyProfessor.com into different groups and apply several LDA models to understand the topics of the comments. Furthermore, we perform sentiment analysis using LIWC2015. We discover a number of  interesting findings that may be helpful for improving the college learning experience for all partied involved, including students, professors and administrators. 

There are three possible directions for future work. First, we can investigate a fine-grained  partition strategy to divide the data by departments, subjects or courses. Second, we can track the comments over time. Our current dataset contain comments from 2018 to 2020, while most of the comments are posted in May and December, which are the ends of spring and fall semesters. With more data over time, we may study individual professor’s teaching style changes and look at this problem from a brand new point of temporal view. Lastly, many in-person lectures are switched to online lectures due to COVID-19 and quarantine. A valuable study is to first determine the courses that are transformed from in-person to online and then understand the changes from the student's experiences.

\bibliography{references}

\begin{thebibliography}{18}
\providecommand{\natexlab}[1]{#1}
\providecommand{\url}[1]{\texttt{#1}}
\providecommand{\urlprefix}{URL }
\expandafter\ifx\csname urlstyle\endcsname\relax
  \providecommand{\doi}[1]{doi:\discretionary{}{}{}#1}\else
  \providecommand{\doi}{doi:\discretionary{}{}{}\begingroup
  \urlstyle{rm}\Url}\fi

\bibitem[{Rat(2020)}]{RateMyProfessors}
 2020.
\newblock Rate My Professors About Page.
\newblock \urlprefix\url{https://www.ratemyprofessors.com/About.jsp}.

\bibitem[{Abulaish et~al.(2009)Abulaish, Jahiruddin, Doja, and
  Ahmad}]{10.1007/978-3-642-11164-8_35}
Abulaish, M.; Jahiruddin; Doja, M.~N.; and Ahmad, T. 2009.
\newblock Feature and Opinion Mining for Customer Review Summarization.
\newblock In Chaudhury, S.; Mitra, S.; Murthy, C.~A.; Sastry, P.~S.; and Pal,
  S.~K., eds., \emph{Pattern Recognition and Machine Intelligence}, 219--224.
  Berlin, Heidelberg: Springer Berlin Heidelberg.
\newblock ISBN 978-3-642-11164-8.

\bibitem[{Arbaugh(2001)}]{doi:10.1177/108056990106400405}
Arbaugh, J. 2001.
\newblock How Instructor Immediacy Behaviors Affect Student Satisfaction and
  Learning in Web-Based Courses.
\newblock \emph{Business Communication Quarterly} 64(4): 42--54.

\bibitem[{Blei, Ng, and Jordan(2003)}]{articleB}
Blei, D.; Ng, A.; and Jordan, M. 2003.
\newblock Latent Dirichlet Allocation.
\newblock \emph{Journal of Machine Learning Research} 3: 993--1022.

\bibitem[{Blei and Lafferty(2006)}]{10.1145/1143844.1143859}
Blei, D.~M.; and Lafferty, J.~D. 2006.
\newblock Dynamic Topic Models.
\newblock In \emph{Proceedings of the 23rd International Conference on Machine
  Learning}, ICML '06, 113–120. New York, NY, USA: Association for Computing
  Machinery.
\newblock ISBN 1595933832.

\bibitem[{Chen et~al.(2020)Chen, Lyu, Yang, Wang, and Luo}]{chen2020eyes}
Chen, L.; Lyu, H.; Yang, T.; Wang, Y.; and Luo, J. 2020.
\newblock In the Eyes of the Beholder: Analyzing Social Media Use of Neutral
  and Controversial Terms for COVID-19.

\bibitem[{Deresiewicz(2014)}]{deresiewicz_2014}
Deresiewicz, W. 2014.
\newblock Don't Send Your Kid to the Ivy League.
\newblock
  \urlprefix\url{https://monticellocollege.org/FileAssets/liberal-arts/ivy_league_schools_are_overrated._send_your_kids_elsewhere.pdf}.

\bibitem[{Etzioni et~al.(2005)Etzioni, Cafarella, Downey, Popescu, Shaked,
  Soderland, Weld, and Yates}]{10.5555/1090483.1644538}
Etzioni, O.; Cafarella, M.; Downey, D.; Popescu, A.-M.; Shaked, T.; Soderland,
  S.; Weld, D.~S.; and Yates, A. 2005.
\newblock Unsupervised Named-Entity Extraction from the Web: An Experimental
  Study.
\newblock \emph{Artif. Intell.} 165(1): 91–134.
\newblock ISSN 0004-3702.

\bibitem[{Hu and Liu(2004)}]{10.1145/1014052.1014073}
Hu, M.; and Liu, B. 2004.
\newblock Mining and Summarizing Customer Reviews.
\newblock KDD '04, 168–177. New York, NY, USA: Association for Computing
  Machinery.
\newblock ISBN 1581138881.

\bibitem[{Miller et~al.(1990)Miller, Beckwith, Fellbaum, Gross, and
  Miller}]{10.1093/ijl/3.4.235}
Miller, G.~A.; Beckwith, R.; Fellbaum, C.; Gross, D.; and Miller, K.~J. 1990.
\newblock {Introduction to WordNet: An On-line Lexical Database*}.
\newblock \emph{International Journal of Lexicography} 3(4): 235--244.
\newblock ISSN 0950-3846.

\bibitem[{Monks and Ehrenberg(1999)}]{NBERw7227}
Monks, J.; and Ehrenberg, R.~G. 1999.
\newblock The Impact of US News and World Report College Rankings on Admission
  Outcomes and Pricing Decisions at Selective Private Institutions.
\newblock Working Paper 7227, National Bureau of Economic Research.

\bibitem[{Morse and Brooks(2020)}]{morse_brooks_2020}
Morse, R.; and Brooks, E. 2020.
\newblock How U.S. News Calculated the 2021 Best Colleges Rankings.
\newblock
  \urlprefix\url{https://www.usnews.com/education/best-colleges/articles/how-us-news-calculated-the-rankings}.

\bibitem[{Newman et~al.(2009)Newman, Asuncion, Smyth, and
  Welling}]{10.5555/1577069.1755845}
Newman, D.; Asuncion, A.; Smyth, P.; and Welling, M. 2009.
\newblock Distributed Algorithms for Topic Models.
\newblock \emph{J. Mach. Learn. Res.} 10: 1801–1828.
\newblock ISSN 1532-4435.

\bibitem[{Otto, Jr, and Ross(2008)}]{doi:10.1080/02602930701293405}
Otto, J.; Jr, D. A.~S.; and Ross, D.~N. 2008.
\newblock Does ratemyprofessor.com really rate my professor?
\newblock \emph{Assessment \& Evaluation in Higher Education} 33(4): 355--368.

\bibitem[{Otto, Sanford, and Wagner(2011)}]{articleA}
Otto, J.; Sanford, D.; and Wagner, W. 2011.
\newblock Analysis Of Online Student Ratings Of University Faculty.
\newblock \emph{Journal of College Teaching \& Learning (TLC)} 2: 25--30.

\bibitem[{Pennebaker et~al.(2015)Pennebaker, Boyd, Jordan, and
  Blackburn}]{pennebaker_boyd_jordan_blackburn_2015}
Pennebaker, J.; Boyd, R.; Jordan, K.; and Blackburn, K. 2015.
\newblock The Development and Psychometric Properties of LIWC2015.
\newblock \emph{University of Texas Libraries}
  \urlprefix\url{http://hdl.handle.net/2152/31333}.

\bibitem[{Silva et~al.(2008)Silva, Silva, Quinn, Draper, Cover, and
  Munoff}]{doi:10.1080/00986280801978434}
Silva, K.~M.; Silva, F.~J.; Quinn, M.~A.; Draper, J.~N.; Cover, K.~R.; and
  Munoff, A.~A. 2008.
\newblock Rate My Professor: Online Evaluations of Psychology Instructors.
\newblock \emph{Teaching of Psychology} 35(2): 71--80.

\bibitem[{Titov and McDonald(2008)}]{titov2008modeling}
Titov, I.; and McDonald, R. 2008.
\newblock Modeling Online Reviews with Multi-grain Topic Models.

\end{thebibliography}

\nocite{titov2008modeling}
\nocite{articleB}
\nocite{10.5555/1577069.1755845}
\nocite{10.1145/1143844.1143859}
\nocite{10.1007/978-3-642-11164-8_35}
\nocite{doi:10.1177/108056990106400405}
\nocite{10.1093/ijl/3.4.235}
\nocite{10.1145/1014052.1014073}
\nocite{10.5555/1090483.1644538}
\nocite{doi:10.1080/00986280801978434}

\end{document}